\pdfoutput=1
\documentclass[11pt,twoside,a4paper,cmspaper,final,collab]{cms-tdr}
\def\svnVersion{fc85396}\def\svnDate{2023/06/29}\def\cmsCernNoTag{CERN-EP-2022-181}\def\cmsCernDate{\today}\def\cmsMessage{Published in Physics Letters B as \href{http://dx.doi.org/10.1016/j.physletb.2023.137803}{\doi{10.1016/j.physletb.2023.137803}.}}
\begin{document}\cmsNoteHeader{EXO-20-011}

\newcommand{\hcmnl}{\ensuremath{\PN_{\ell}}\xspace}
\newcommand{\hcmne}{\ensuremath{\PN_{\Pe}}\xspace}
\newcommand{\hcmnmu}{\ensuremath{\PN_{\PGm}}\xspace}
\newcommand{\mhcmn}{\ensuremath{m_{\hcmnl}}\xspace}
\newcommand{\wjets}{{{\PW}\!+jets}\xspace}
\newcommand{\qqbarp}{\ensuremath{\PQq\PAQq^\prime}\xspace}
\newcommand{\eeqq}{\ensuremath{\Pe\Pe\qqbarp}\xspace}
\newcommand{\mmqq}{\ensuremath{\PGm\PGm\qqbarp}\xspace}
\newcommand{\mllj}{\ensuremath{m(\ell\ell\,\mathrm{J})}\xspace}
\newcommand{\mll}{\ensuremath{m(\ell\ell)}\xspace}
\newcommand{\memuj}{\ensuremath{m(\Pe\PGm \mathrm{J})}\xspace}
\newcommand{\sigstr}{\ensuremath{\mu}\xspace}
\newlength\cmsTabSkip\setlength{\cmsTabSkip}{1ex}
\ifthenelse{\boolean{cms@external}}{\providecommand{\cmsTable}[1]{\resizebox{0.49\textwidth}{!}{#1}}}{\providecommand{\cmsTable}[1]{#1}} 
\ifthenelse{\boolean{cms@external}}{\providecommand{\cmsLeft}{upper\xspace}}{\providecommand{\cmsLeft}{left\xspace}}
\ifthenelse{\boolean{cms@external}}{\providecommand{\cmsRight}{lower\xspace}}{\providecommand{\cmsRight}{right\xspace}}

\cmsNoteHeader{EXO-20-011}
\title{Search for a heavy composite Majorana neutrino in events with dilepton signatures from proton-proton collisions at \texorpdfstring{$\sqrt{s} = 13\TeV$}{sqrt(s) = 13 TeV}}

\date{\today}

\abstract{
Results are presented of a search for a heavy Majorana neutrino \hcmnl decaying into two same-flavor leptons $\ell$ (electrons or muons) and a quark-pair jet. A model is considered in which the \hcmnl is an excited neutrino in a compositeness scenario. The analysis is performed using a sample of proton-proton collisions at $\sqrt{s} = 13\TeV$ recorded by the CMS experiment at the CERN LHC, corresponding to an integrated luminosity of 138\fbinv. The data are found to be in agreement with the standard model prediction. For the process in which the \hcmnl is produced in association with a lepton, followed by the decay of the \hcmnl to a same-flavor lepton and a quark pair, an upper limit at 95\% confidence level on the product of the cross section and branching fraction is obtained as a function of the \hcmnl mass \mhcmn and the compositeness scale $\Lambda$. For this model the data exclude the existence of \hcmne (\hcmnmu) for \mhcmn below 6.0 (6.1)\TeV, at the limit where \mhcmn is equal to $\Lambda$. For $\mhcmn \approx 1\TeV$, values of $\Lambda$ less than 20 (23)\TeV are excluded.
These results represent a considerable improvement in sensitivity, covering a larger parameter space than previous searches in $\Pp\Pp$ collisions at 13\TeV.
}

\hypersetup{
pdfauthor={CMS Collaboration},
pdftitle={Search for a heavy composite Majorana neutrino in events with dilepton signatures from proton-proton collisions at sqrt(s) = 13 TeV},
pdfsubject={CMS},
pdfkeywords={CMS, LHC, composite fermions, heavy leptons}
}

\maketitle

\section{Introduction}
\label{sec:introduction}

The standard model (SM) of particle physics is an extremely successful theory that has been extensively verified against experimental results. 
Nevertheless, there are several fundamental aspects of particle phenomenology that are not explained within the SM.
One of these is the appearance of three generations of leptons and quarks, regarded as fundamental fermions in the SM, and the related question of the mass hierarchy across the generations. A possible solution to these issues is offered by composite-fermion models \cite{{Pati:1975md}, {Pati:1983dh},{Harari:1982xy},{Greenberg:1974qb},{Dirac:1963aa},{Terazawa:1976xx},{Eichten:1979ah},{Eichten:1983hw},{Baur:1989kv},{Baur:1987ga}}, in which the quarks and leptons have substructure.

In the composite-fermion scenario, quarks and leptons are assumed to have an internal substructure that would manifest itself at some sufficiently high energy scale $\Lambda$, the compositeness scale. This scale plays the role of an expansion parameter with which a series of higher-dimensional operators are constructed in an effective field theory (EFT) framework. The fermions of the SM are considered as bound states of some not-yet-observed fundamental constituents, generically referred to as \emph{preons} \cite{Pati:1983dh}.
Two model-independent features~\cite{Terazawa,Eichten:1983hw,Cabibbo,Baur:1989kv} are experimentally relevant: excited states of quarks and leptons with masses lower than or equal to $\Lambda$, and gauge or contact effective interactions (GI or CI) between the ordinary fermions and these excited states. The gauge interaction involves both fermion and gauge boson fields, and, at the lowest order in the EFT expansion, is described by dimension-five operators. Conversely, the contact interaction involves only fermion fields, with corresponding operators of dimension six.

A particular case of such excited states is a heavy composite Majorana neutrino (\hcmnl, $\ell = \Pe,\,\PGm,\,\PGt$)~\cite{O3,O2,O1,Biondini:2021vip}, a neutral lepton having a mass above the electroweak energy scale.
The introduction of an \hcmnl is well motivated as an explanation of the baryon asymmetry in the universe.
Indeed, in the framework of baryogenesis via leptogenesis~\cite{Fukugita:1986hr,Davidson:2008bu}, heavy Majorana fermions are the source of the matter-antimatter asymmetry in CP violating decays in the early universe, and it has been proposed \cite{Zhuridov:2016xls,Biondini:2017fut} that \hcmnl's could quantitatively account for the observed asymmetry. Such composite Majorana neutrinos would also lead to observable effects in neutrinoless double beta decay experiments \cite{O2,Biondini:2021vip}.

\begin{figure*}[htb]
\centering
\includegraphics[width=\textwidth]{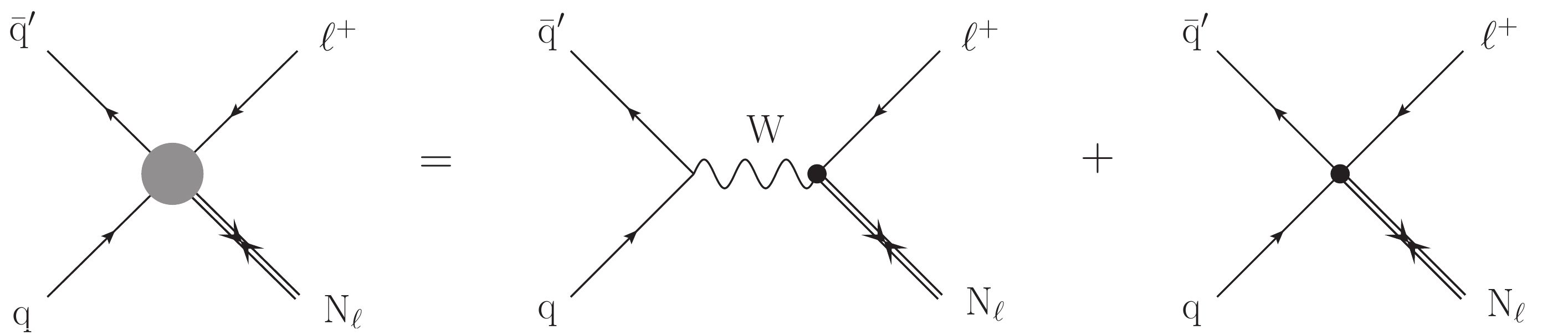}
\caption{The fermion interaction as a sum of gauge (center) and contact (right) contributions.}
\label{interactions}
\end{figure*}

\begin{figure*}[htb]
\centering
\includegraphics[width=\textwidth]{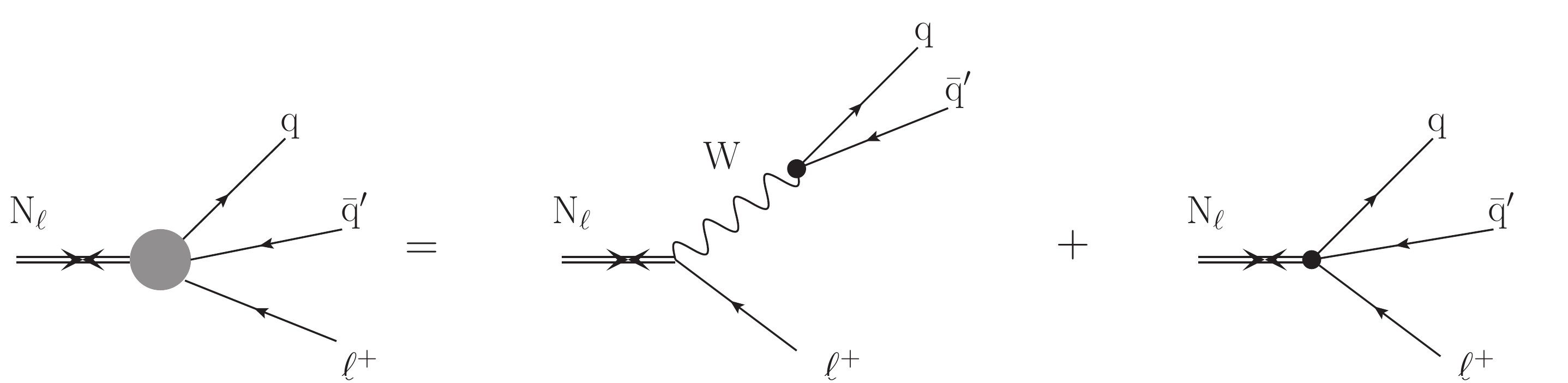}
\caption{Feynman diagrams for the decay of a heavy composite Majorana neutrino to $\ell\qqbarp$.}
\label{decay}
\end{figure*}

As a general phenomenological framework we consider the composite neutrino model given in Ref.~\cite{Leonardi:2015qna}, in which the GI and CI enter into both the production and decay of \hcmnl's and are governed, respectively, by the effective Lagrangians
\begin{linenomath}
\begin{align}
    \mathcal{L}_{\text{GI}} &= \frac{g \, f}{\sqrt{2} \Lambda} \overline{N} \sigma^{\mu \nu} (\partial_\mu W_\nu) P_{\text{L}} \ell + \text{h.c.},
    \label{GI_Lag_LH}\\
    \mathcal{L}_{\text{CI}} &= \frac{g_*^2 \, \eta}{\Lambda^2} \bar{q}^\prime \gamma^\mu P_{\text{L}} q \, \overline{N} \gamma_\mu P_{\text{L}} \ell  + \text{h.c.}
    \label{CI_Lag_LH}
\end{align}
\end{linenomath}
Here $N$, $\ell$, $W$, and $q$ are the \hcmnl, charged lepton, \PW boson, and quark fields, respectively, $P_{\text{L}}$ is the left-handed chirality projection operator, and $g$ is the $\text{SU(2)}_{\text{L}}$ gauge coupling.  The effective coupling for contact interactions, $g^2_*$, takes the value $4\pi$~\cite{Leonardi:2015qna}. The factors $f$ and $\eta$ are additional couplings in the composite model; they are taken here to be unity, a choice that is commonly adopted in phenomenological studies and experimental analyses of composite-fermion models. 
The total amplitude for the production process is given by the coherent sum of the gauge and contact contributions, as shown in Fig.~\ref{interactions}, as well as for the decay modes shown in Fig.~\ref{decay}. 
The production cross section via contact interaction is dominant for a wide range of $\Lambda$ values, including the ones to which this search is sensitive. 

In this work, we consider a composite neutrino, produced in association with a charged lepton, that subsequently decays to a charged lepton and a pair of quarks, leading to the experimental signature $\ell\ell\qqbarp$.  
Because the \hcmnl is a Majorana lepton at the TeV scale, the expected signal is characterized by two leptons $\ell$ with high transverse momentum (\pt) that may be of the same or opposite charge sign, but are of the same flavor. 
We focus on the cases in which these leptons are both electrons or both muons, and the quark pair is detected as a wide jet.
A shape-based analysis is performed, searching for evidence of a signal in the distribution of the invariant mass of the system comprising the two leptons and the quark-pair jet.

The data sample of proton-proton collisions at $\sqrt{s} = 13\TeV$ was recorded in 2016--2018 with the CMS detector at the CERN LHC, and corresponds to an integrated luminosity of 138\fbinv.
A previous search for \hcmnl was performed by CMS with a data sample corresponding to 2.3\fbinv at $\sqrt{s} = 13\TeV$~\cite{2017315}, and found agreement between the data and SM expectations. A 95\% confidence level~(\CL) upper limit on the Majorana neutrino mass \mhcmn was placed at about 4.6\TeV for both the electron and muon channels.
With the larger statistical power of the current data sample, the present search explores a wider range of the parameter space (\mhcmn, $\Lambda$).
We further expand the composite model with recent considerations on the scope of validity of the effective operators in Eqs.~(\ref{GI_Lag_LH}) and (\ref{CI_Lag_LH}) as derived in Ref.~\cite{Biondini:2019tcc}. The unitarity bounds on these operators are used as guidance to optimize the search and extend the analysis sensitivity to lower \mhcmn and higher $\Lambda$ compared with the previous search. 
Tabulated results are provided in the HEPData record for this analysis~\cite{hepdata}.  

More generally, excited states interacting with the SM sector have been extensively searched for at high-energy collider facilities. The current most stringent bounds come from the recent LHC experiments. Excited charged leptons ($\Pe^*,\PGm^*$) have been searched for in the channel $\Pp\Pp \to \ell\ell^* \to \ell\ell\PGg$~\cite{atlas-limit,atlas-limit-new,cms-limit-7TeV,cms-limit-8TeV,CMS:2018sfq,CMS:2020cay}, where they would be produced via CI and then decay via GI, and in the channel $\Pp\Pp \to \ell\ell^* \to \ell\ell\qqbarp$~\cite{CMS:2020cay} where both production and decay proceed through CI.

\section{The CMS detector}
\label{sec:detector}
The central feature of the CMS apparatus is a superconducting solenoid, of 6\unit{m} internal diameter, providing a field of 3.8\unit{T}. 
Within the field volume, there are the inner tracker, the crystal electromagnetic calorimeter (ECAL), and the brass and scintillator hadron calorimeter (HCAL).
The inner tracker is composed of a pixel detector and a silicon strip tracker, and measures charged-particle trajectories
in the pseudorapidity range $\abs{\eta} < 2.5$.
The finely segmented ECAL consists of nearly $76\,000$ lead-tungstate crystals that provide coverage up to $\abs{\eta} = 3.0$.
The HCAL consists of a sampling calorimeter, which utilizes alternating layers of brass as an absorber and
plastic scintillator as an active material, covering the range $\abs{\eta} < 3$, and is extended to $\abs{\eta} < 5$ by the forward hadron calorimeters.
The muon system covers the region $\abs{\eta} < 2.4$ and consists of up to four planes of gas ionization muon detectors installed outside the solenoid and sandwiched between the layers of the steel flux-return yoke.
Events of interest are selected using a two-tiered trigger system. The first level, composed of custom hardware processors, uses information from the calorimeters and muon detectors to select events at a rate of around 100\unit{kHz} within a fixed latency of about 4\mus~\cite{CMS:2020cmk}. The second level, known as the high-level trigger, consists of a farm of processors running a version of the full event reconstruction software optimized for fast processing, and reduces the event rate to around 1\unit{kHz} before data storage~\cite{CMS:2016ngn}.
A detailed description of the CMS detector can be found in Ref.~\cite{Chatrchyan:2008aa}.

\section{Monte Carlo simulation}
\label{sec:simulation}

The signal and the SM backgrounds are simulated using the Monte Carlo (MC) method.
The simulated samples for the signal are generated at leading order (LO) with \CALCHEP v3.6~\cite{Belyaev:2012qa}, using the NNPDF~3.0 LO parton distribution functions (PDFs) with the four-flavor scheme~\cite{Ball:2014uwa}. Samples are generated for $\Lambda$ values from 4 to 20\TeV, and with \mhcmn values from 0.5\TeV to $\Lambda$, the maximum value consistent with the model.

The background processes simulated are top quark pair production \ttbar, single top quark production $\PQt\PW$, the Drell--Yan (DY) process, \wjets, diboson production ($\PW\PW$, $\PW\PZ$, $\PZ\PZ$), \ttbar with vector boson production $\ttbar\textrm{V}$, and SM production of jets through the strong interaction described by quantum chromodynamics (QCD).
The \ttbar events are generated at next-to-leading order (NLO) with \POWHEG~v2.0~\cite{Nason:2004rx,Frixione:2007vw,Alioli:2010xd,Alioli:2009je,Re:2010bp}. 
The \POWHEG generator is also used to describe $\PQt\PW$ production at NLO.  
The DY, QCD, \wjets, and $\ttbar\textrm{V}$ samples are generated at LO with \MGvATNLO~v2.2.2 (v2.4.2)~\cite{Alwall:2014hca} for the 2016 (2017--2018) samples.
The DY events are weighted by a \pt-dependent $K$ factor, a function of the generator-level \PZ boson momentum $\pt(\PZ)$. The $K$ factor serves both to adjust a mismodeling of the $\pt(\PZ)$ distribution~\cite{Lindert:2017olm,CMS:2019raw} and to account for higher-order effects in the QCD and EW perturbative expansions. It is a product of two terms: one obtained as described in Ref.~\cite{CMS:2017zts} from Drell--Yan NLO samples, produced with \MADGRAPH{}5\_a\MCATNLO with the FXFX matching scheme~\cite{Frederix:2012ps}, while the other is extracted from theoretical calculations~\cite{Lindert:2017olm}.
The diboson processes are generated with \PYTHIA~\cite{Sjostrand:2014zea} at LO.

For the simulation of all backgrounds we use the {NNPDF}~3.0~\cite{Ball:2014uwa} PDFs for 2016, and {NNPDF}~3.1 next-to-NLO (NNLO)~\cite{NNPDF:2017mvq} for 2017--2018 samples. 
Parton showering and hadronization are described by \PYTHIA~8.226 (8.230) with the {CUETP8M1}~\cite{CMS:2015wcf} ({CP5}~\cite{CMS:2019csb}) tune for 2016 (2017--2018) samples.
Additional collisions in the same or adjacent bunch crossings (pileup) are taken into account
by superimposing simulated minimum bias interactions onto the hard scattering process,
with a number distribution matching that observed in data. Simulated events are propagated through
the \GEANTfour~\cite{Agostinelli:2002hh} based simulation of the CMS detector, tuned for detector-related differences in each data-taking period.
Normalization of the simulated background samples is performed using the most precise cross section calculations available~\cite{Alioli:2009je,Re:2010bp,Alwall:2014hca,Melia:2011tj,Beneke:2011mq,Cacciari:2011hy,Baernreuther:2012ws,Czakon:2012zr,Czakon:2012pz,Czakon:2013goa,Gavin:2010az}, which are generally calculated to NLO or NNLO.

\section{Event and object selection}
\label{sec:evtSel}
Single-lepton triggers that require either an electron with $\pt > 115\GeV$ within $\abs{\eta} < 2.5$ or a muon with $\pt > 50\GeV$ within $\abs{\eta} < 2.4$ are used to select events in the $\Pe\Pe\qqbarp$ and $\PGm\PGm\qqbarp$ channels, respectively.
The separate \pt requirements reflect different trigger thresholds; these do not affect the relative signal sensitivity, as the signal is characterized by high-momentum leptons in the final state.
The primary vertex (PV) of the event is taken to be the vertex corresponding to the hardest scattering in the event, evaluated using tracking information alone, as described in Section 9.4.1 of Ref.~\cite{CMS-TDR-15-02}.
Electrons are reconstructed as superclusters in the ECAL associated with tracks in the tracking detector~\cite{CMS:2020uim,CMS-DP-2020-021}.
Requirements on energy deposits in the calorimeter and the number of track measurements are imposed to distinguish electrons from charged pions, and
electrons associated with the PV from those produced by photon conversions.
Muons are reconstructed using the tracker and muon detectors.
Quality requirements, based on the minimum number of measurements in the silicon tracker, pixel detector, and muon detectors are applied to suppress backgrounds from hadron decays displaced from the PV and
from hadron shower remnants that reach the muon system~\cite{CMS:2019ied}.
We require exactly two electrons, or two muons, that originate from the PV.

The \pt of the leading (subleading) lepton is required to be higher than 150 (100)\GeV.
Isolation requirements are imposed to suppress backgrounds from jets that are misidentified as leptons or that contain leptons from heavy-flavor hadron decays.
The isolation is defined as the \pt sum of tracks within a cone around the candidate direction of size $\Delta R \equiv \sqrt{\smash[b]{(\Delta\eta)^2+(\Delta\phi)^2}} = 0.3$, where $\phi$ is the azimuthal angle in radians. The momentum of the candidate is excluded from the sum. The isolation is required to be less than 3 (10)\% of candidate electron (muon) \pt.

Jets are reconstructed using the anti-\kt clustering scheme~\cite{Cacciari:2008gp,Cacciari:2011ma}
applied to the objects reconstructed with a particle-flow algorithm~\cite{CMS-PRF-14-001}.
The latter combines information from all CMS subdetectors and reconstructs individual particles in the event (electrons, muons, photons, and neutral and charged hadrons).
Jets are reconstructed with a distance parameter $R = 0.8$ and are referred to here as ``large-radius jets'', labeled by the symbol ``J''. This value of $R$ is chosen to capture both final-state quarks as a single jet.
The large-radius jets are required to have $\pt > 190\GeV$, $\abs{\eta} < 2.4$, and to be separated from leptons
by $\Delta R > 0.8$.
The pileup per particle identification algorithm (PUPPI)~\cite{Sirunyan:2020foa,Bertolini:2014bba} is used to mitigate the effect of pileup at the reconstructed particle level,
making use of event pileup properties, tracking information, and a local shape variable that distinguishes between collinear and soft diffuse distributions of other particles surrounding the particle under consideration. The collinear component is attributed to particles originating from the hard scatter, and the soft diffuse one to particles originating from pileup interactions.

\section{Analysis strategy and background estimation}
\label{sec:background}

To define the signal region (SR) for the search we require that the event contain two same-flavor leptons with invariant mass $\mll > 300\GeV$, together with at least one large-radius jet.
The requirement on \mll is introduced to reduce the DY background and part of the \ttbar background, with minimal effect on the signal acceptance. 
No requirement is placed on the charge of the leptons, to retain efficiency for both same and opposite sign signal events and to avoid the systematic uncertainty associated with the efficiency of the charge sign determination for such energetic particles.
While a veto of opposite-sign lepton pairs would reduce the SM background, optimization studies have shown that it is better to impose kinematical requirements that retain the signal efficiency at high momenta.

For gauge-mediated decays of the \hcmnl, the fragmentation products of the two quarks from the \PW boson decay typically form at least one large-radius jet. In the case of contact-mediated decays, the two quarks are well separated, but at least one of them will be contained within a large-radius jet.
The signal simulation shows that the efficiency for capturing one or both quarks in the jet is 98\% for the CI-dominated case $\mhcmn = 5\TeV$, and 95\% overall for the gauge or contact interaction with $\mhcmn > 1\TeV$.

The key variable for the analysis is the invariant mass \mllj of the system comprising the two leptons and the leading large-radius jet. This variable provides good discrimination between the signal and SM background contributions and is also correlated with \mhcmn, which would become relevant for the signal characterization if an excess were observed. 
The statistical analysis is implemented with a maximum likelihood (ML) fit to extract the signal strength \sigstr, the ratio of the signal yield observed to that predicted by the model. The inputs to the fit are the distributions in \mllj of the estimated backgrounds, the expected signal, and the data.
 
The leading background to this search is DY production of a lepton pair accompanied by a jet from initial-state radiation. The second major background comes from processes that produce top quarks, \ttbar and $\PQt\PW$.

The DY contribution is estimated from simulation, corrected by
scale factors that serve to adjust the simulated \mllj shape for differences with respect to the data. A scale factor for each \mllj bin and each data-taking year is taken from the DY-dominated \mll region around the \PZ boson mass peak, $60 < \mll < 120\GeV$.  
The scale factors have values in the range 0.87--1.57.  Their statistical uncertainties are combined with those of the simulation to estimate the total systematic uncertainty in the DY background prediction to be used in the fit.

In addition, we include in the fit an SM-dominated control region (CR) selected with the same criteria as the SR except in an \mll band, $150 < \mll < 300\GeV$, that lies adjacent to the SR.  This CR provides validation of the corrected simulation and improves the precision of the background prediction. 

\begin{figure*}
\centering
\includegraphics[width=0.49\textwidth]{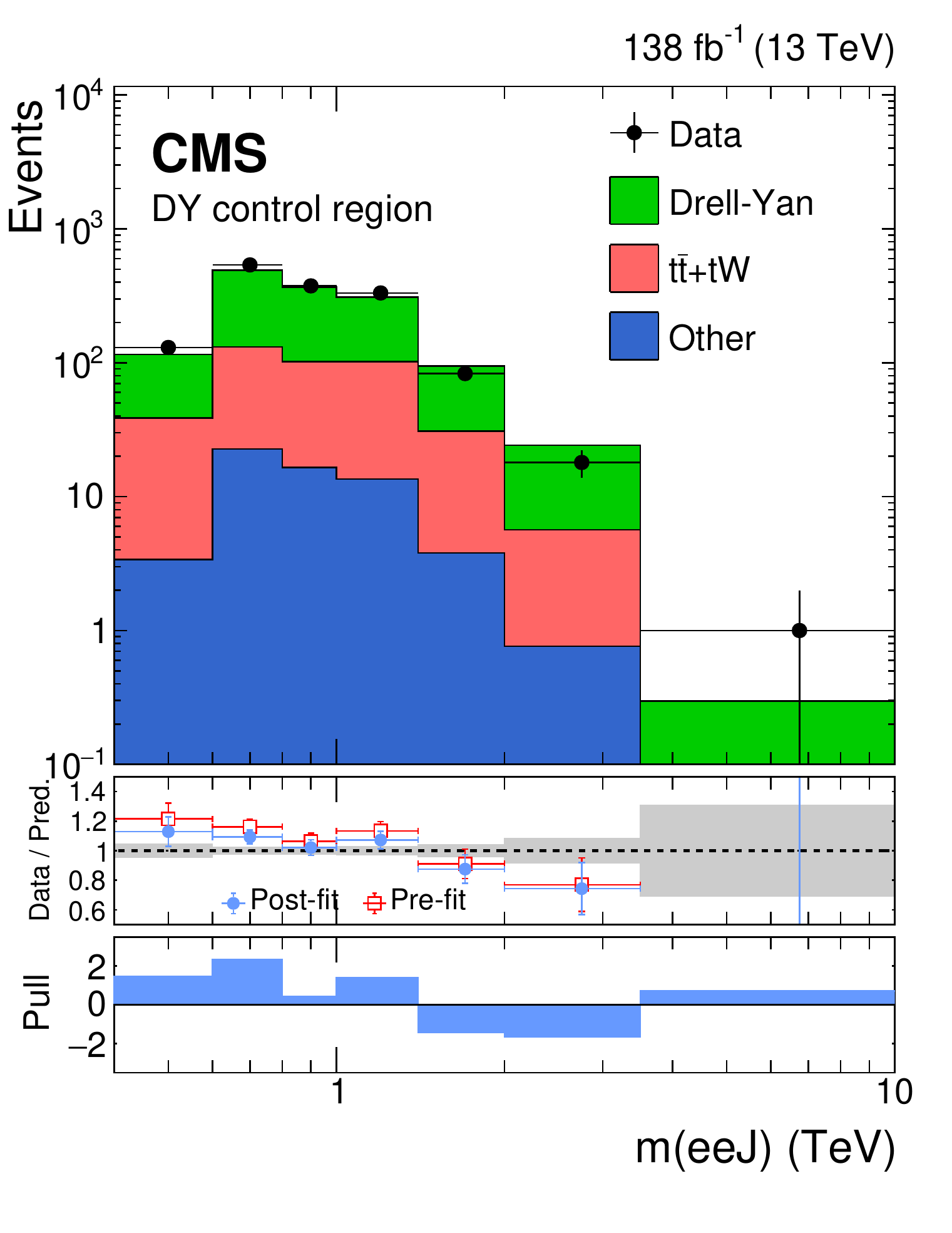}
\includegraphics[width=0.49\textwidth]{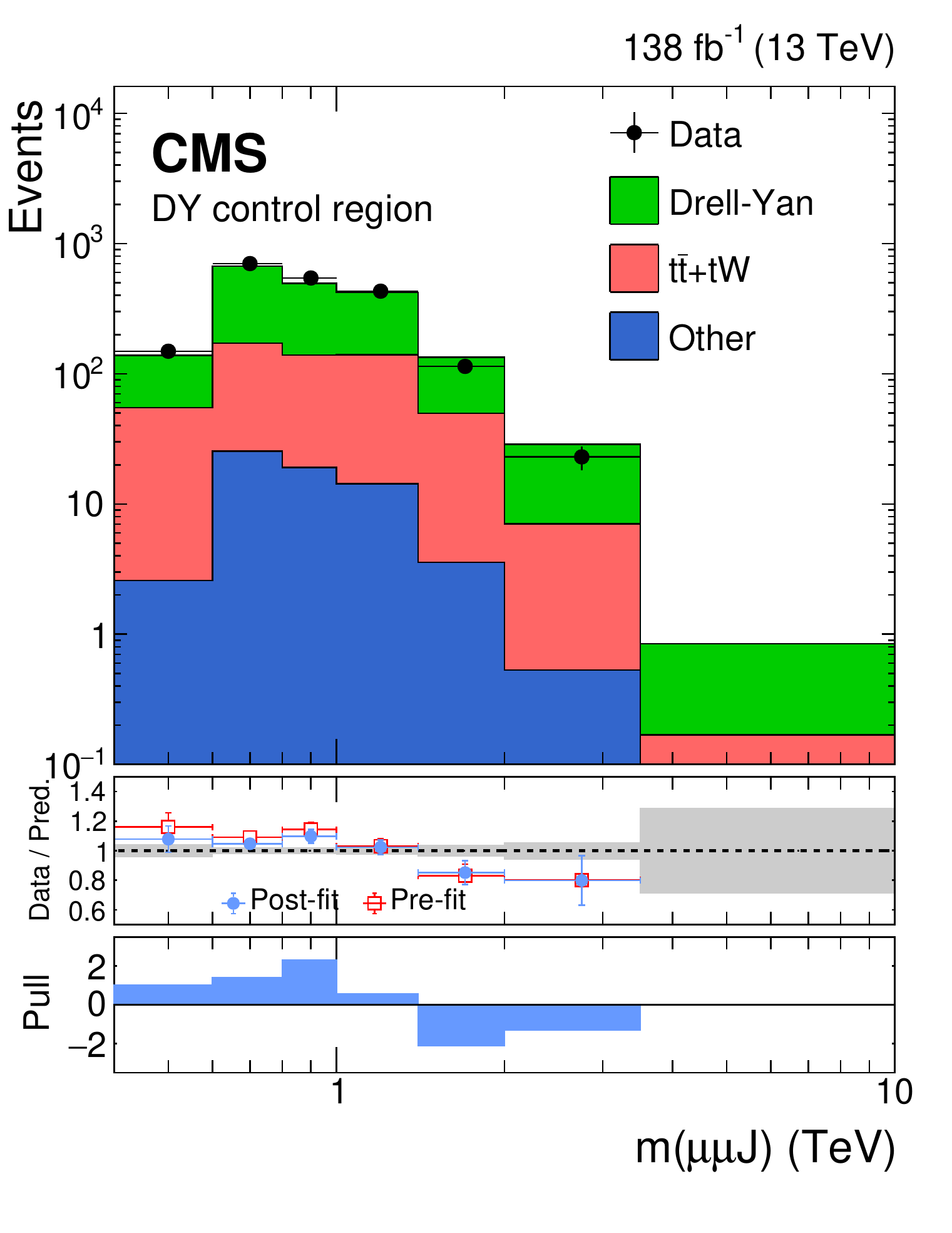}\\
\includegraphics[width=0.49\textwidth]{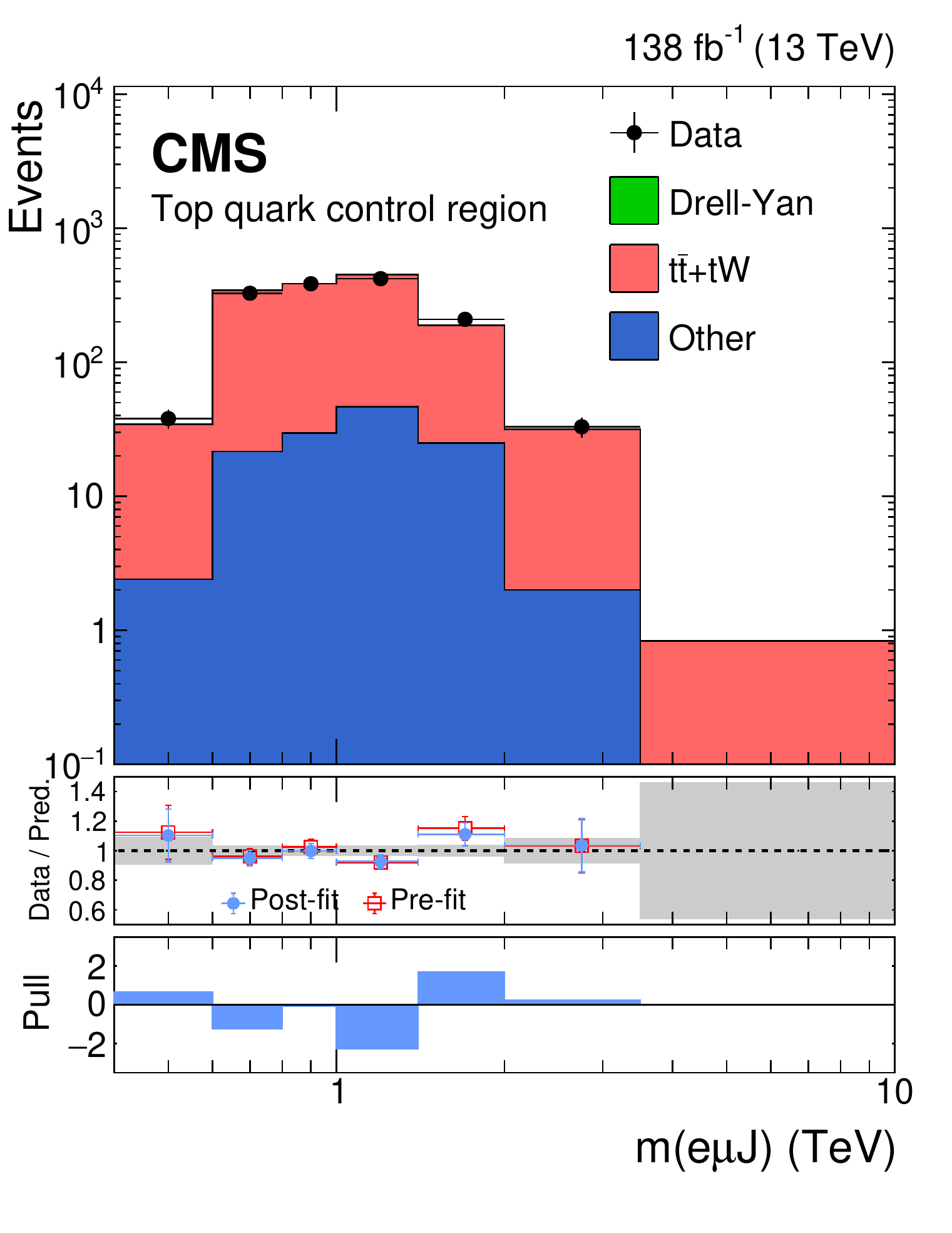}
\caption{Distribution of \mllj in the DY-enriched CR for the electron (upper left) and muon (upper right) flavors, and of \memuj in the top-quark-enriched CR (lower). 
Data points are overlaid on the post-fit background (stacked histograms).
The overflow is included in the last bin.
The middle panels show ratios of the data to the pre-fit background prediction and post-fit background yield as red open squares and blue points, respectively.
The gray band in the middle panels indicates the systematic component of the post-fit uncertainty.
The lower panels show the distributions of the pulls, defined in the text.
} 
\label{fig:bkg CR}
\end{figure*}

The \mllj distributions of the electron and muon DY CRs are shown in Figure~\ref{fig:bkg CR}, upper left and right, respectively.  The distributions for both flavors are included in the ML fit to constrain the DY background contribution. In the figure the data are compared with the background estimated before (pre-fit) and after (post-fit) the simultaneous fit of the signal and control regions.
The pulls shown in the lower panels are defined as the difference between data and the post-fit background prediction divided by the quadratic difference of the uncertainties in the data and the post-fit yields. The quadratic difference of the uncertainties is taken to account for the correlation between the data and the post-fit prediction.

The second most important background arises from the leptonic decays of top quarks from \ttbar and single top quark production. The \mllj shape of this background is taken from the MC simulation, with a free normalization parameter in the ML fit for each data-taking year.  To constrain these parameters, we include a top quark enriched CR in the fit.  The definition of the CR exploits
the fact that the decays of the top quarks from \ttbar production give rise to events with $\Pe\Pe$, $\PGm\PGm$, and $\Pe\PGm$ configurations, with the $\Pe\PGm$ final state having a branching fraction twice that of either of the same-flavor pairs.
We thus select events having one muon and one electron,
the leading (subleading) lepton having $\pt > 150\,(100)\GeV$.
In addition, we reject events containing an electron and a muon, each with $\pt > 5\GeV$, that have angular separation $\Delta R < 0.1$.
Figure~\ref{fig:bkg CR} (lower) shows the \memuj distribution of the CR data along with pre- and post-fit background estimates.

The remaining SM backgrounds, arising from QCD, \wjets, and diboson production, are small (${\sim}5\%$ of the total).  Their contribution is taken directly from MC simulation, normalized to the theoretical cross sections cited in Section \ref{sec:simulation}. These processes are designated ``other'' in the legends in Figs.~\ref{fig:bkg CR} and \ref{fig:postfitSR}. In the three highest \mllj bins, which are the most sensitive to a signal, the fractions of backgrounds are approximately 60, 34, and 6\% for DY, \ttbar, and Other, respectively. The total event yield information for Fig.~\ref{fig:postfitSR} is available in the HEPData record for this analysis~\cite{hepdata}.

\section{Systematic uncertainties}
\label{sec:systematics}

Sources of systematic uncertainties affecting the \mllj distribution include statistical uncertainties in the CR data and in the simulation, together with systematic uncertainties in quantities affecting the modeling in the simulation.  The latter are accounted for with log-normal-distributed nuisance parameters in the fitting procedure described in Section~\ref{sec:background}.
Uncertainties from a given source are treated as uncorrelated across the three data-taking years, with the exception of electron energy scale and resolution, small-background theoretical cross sections, and signal shape, which are fully correlated.

The integrated luminosities for the 2016, 2017, and 2018 data-taking years have individual uncertainties of 1.2--2.5\%~\cite{CMS-LUM-17-003,CMS-PAS-LUM-17-004,CMS-PAS-LUM-18-002}, while the overall uncertainty for the 2016--2018 period is 1.6\%.
These uncertainties affect the normalization of signal yields and those background yields that are taken from simulation.
The imperfect modeling of pileup interactions is estimated by varying the total cross section for inelastic $\Pp\Pp$ scattering used in the simulation by ${\pm}5\%$~\cite{CMS:2018mlc}, and results in an uncertainty of 0.004 (0.006) in the fitted value of \sigstr in the electron (muon) channel. 

The lepton trigger, reconstruction, identification, and isolation efficiencies are measured in both data and simulation using $\PZ \to \ell\ell$ events. Data-to-simulation scale factors are applied to all simulation samples to account for the differences observed between the two.
The uncertainties in the lepton scale factors are propagated to the estimation of \sigstr, and their effect is found to be 0.004 (0.006) for the electron (muon) signal.

Similarly, the momenta of leptons are varied in the simulation within their uncertainties from the nominal values to ascertain the effect of these uncertainties on the mass distributions. To evaluate the effect of the uncertainty in the momentum resolution of very high energy muons, a Gaussian smearing is applied to the muon momentum and propagated to the \mllj distribution; the resulting effect on the signal yield is less than 0.3\%, with a negligible impact on \sigstr.
The uncertainties in the jet energy scale and resolutions~\cite{CMS:2016lmd} affect the uncertainty in \sigstr by 0.006--0.010.

For the DY background simulation we account for uncertainties in the higher-order QCD and EW corrections and in the data-to-simulation scale factors. The uncertainty in the $\pt(\PZ)$ reweighting described in Section~\ref{sec:simulation} accounts for theoretical uncertainties, implemented as described in Ref.~\cite{CMS:2017zts}, and a component due to the MC samples, applied by varying the $K$ factor by its upward and downward statistical uncertainty. Similarly, the scale factors are varied within their uncertainties to estimate their impact on the invariant mass distributions. The resulting uncertainties in the signal strength are up to 0.034--0.045 in the two leptonic final states considered. 	

Leaving the normalization of the top quark processes floating in the fit results in an uncertainty in the signal strength of up to 0.009 (0.008) in the electron (muon) channel.
For the smaller SM backgrounds, the uncertainty in the cross section is used. 
The theoretical uncertainties in the signal simulation originating from the PDFs have been computed using the recommendations of Ref.~\cite{Butterworth:2015oua}, extracting weights from the MC replicas that vary from a few percent to about 8\%, depending on \mhcmn. These weights affect the selection efficiency, but their uncertainties make a negligible contribution to the final systematic uncertainty.
Finally, uncertainties related to the limited number of simulated events are taken into account with the Barlow--Beeston lite approach~\cite{Barlow:1993dm}. They are considered for all bins of the distributions that are used to extract the results, and kept uncorrelated across the different samples and across the bins of an individual distribution~\cite{Conway:2011in}. The limited size of the simulated event and data samples are two major sources of uncertainty and account for up to 0.021 and 0.083 in \sigstr for both lepton channels.

The impacts of the systematic uncertainties on \sigstr as extracted from the ML fit are summarized in Table~\ref{tab:systUnc HCN run2}. We have checked the sensitivity of the results to variations of the individual nuisance parameters; no significant overconstraining or underestimating of the systematic uncertainties was found.
\begin{table}[htb]
\centering
\topcaption{The impact of each systematic uncertainty on the signal strength \sigstr as extracted from the ML fit, for the \hcmnl signal point with $\mhcmn = 0.5\TeV$ and $\Lambda = 13\TeV$.  Upper and lower uncertainties are given, for both electron and muon channels.}
\cmsTable{
\begin{tabular}{l ll ll}
\hline
Source & \multicolumn{2}{c}{$\Pe\Pe\qqbarp$} & \multicolumn{2}{c}{$\PGm\PGm\qqbarp$} \\
\hline 
Luminosity 				                  & $+0.004$ & $-0.002$ & $+0.002$ & $-0.004$ \\
Lepton scale factors                & $+0.004$ & $-0.002$ & $+0.006$ & $-0.006$ \\
Jet energy scale and resolution     & $+0.010$ & $-0.006$ & $+0.006$ & $-0.003$ \\
Pileup                              & $+0.004$ & $-0.001$ & $+0.006$ & $-0.003$ \\
Lepton energy scale and resolution  & $+0.021$ & $-0.017$ & $+0.002$ & $-0.002$ \\
\ttbar, $\PQt\PW$ normalization     & $+0.009$ & $-0.006$ & $+0.008$ & $-0.002$ \\
DY scale factors                    & $+0.043$ & $-0.045$ & $+0.033$ & $-0.034$ \\
DY $K$ factor                       & $+0.008$ & $-0.007$ & $+0.007$ & $-0.003$ \\
Theory (PDF, cross section)         & $+0.005$ & $-0.004$ & $+0.002$ & $-0.002$ \\
Limited MC sample size              & $+0.021$ & $-0.019$ & $+0.013$ & $-0.005$ \\[\cmsTabSkip]
Statistical                         & $+0.083$ & $-0.074$ & $+0.061$ & $-0.043$ \\
Total                               & $+0.100$ & $-0.091$ & $+0.072$ & $-0.055$ \\
\hline
\end{tabular}
}
\label{tab:systUnc HCN run2}
\end{table}

\section{Results}
\label{sec:results}
\begin{figure*}[htbp]
\centering
\includegraphics[width=0.49\textwidth]{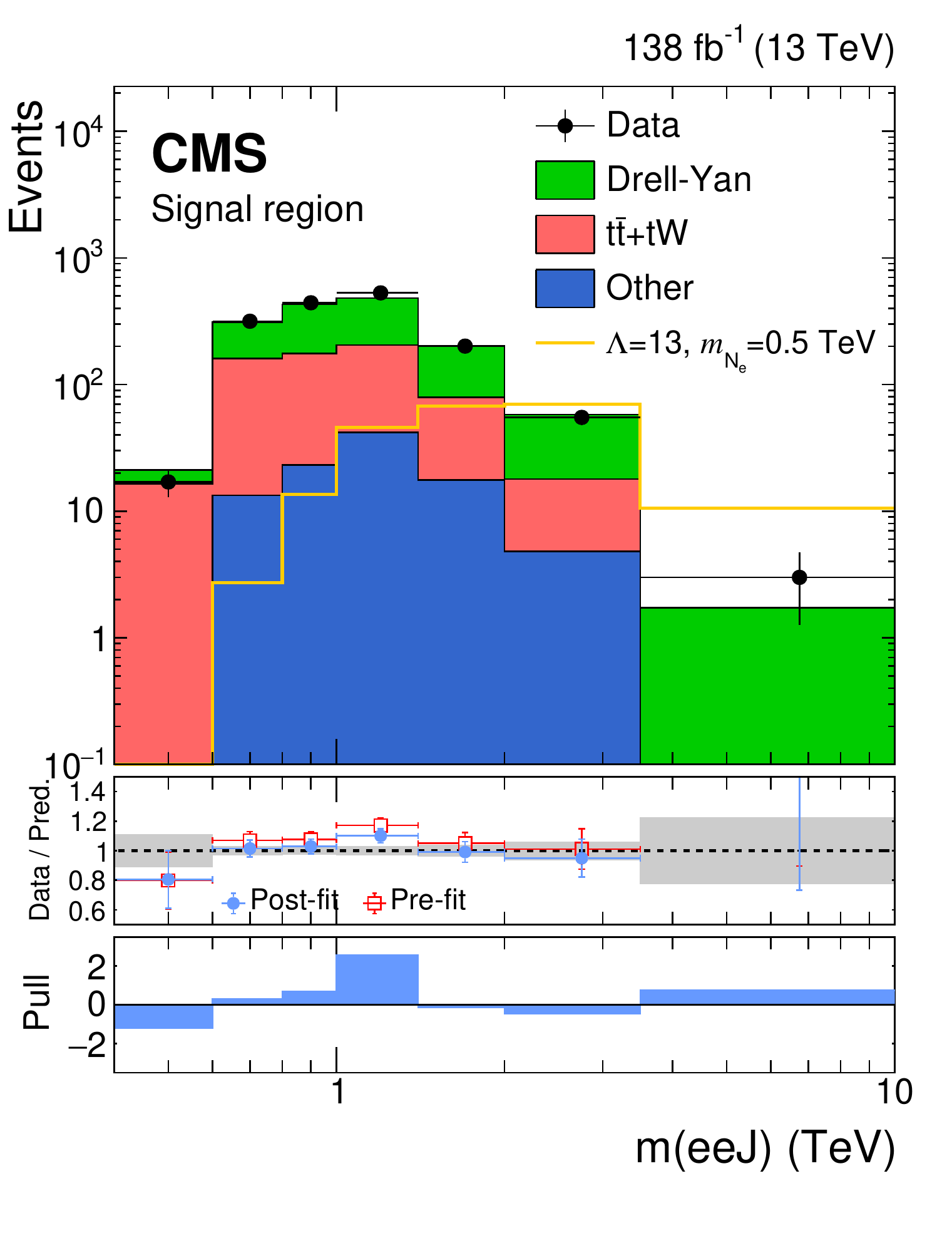}
\includegraphics[width=0.49\textwidth]{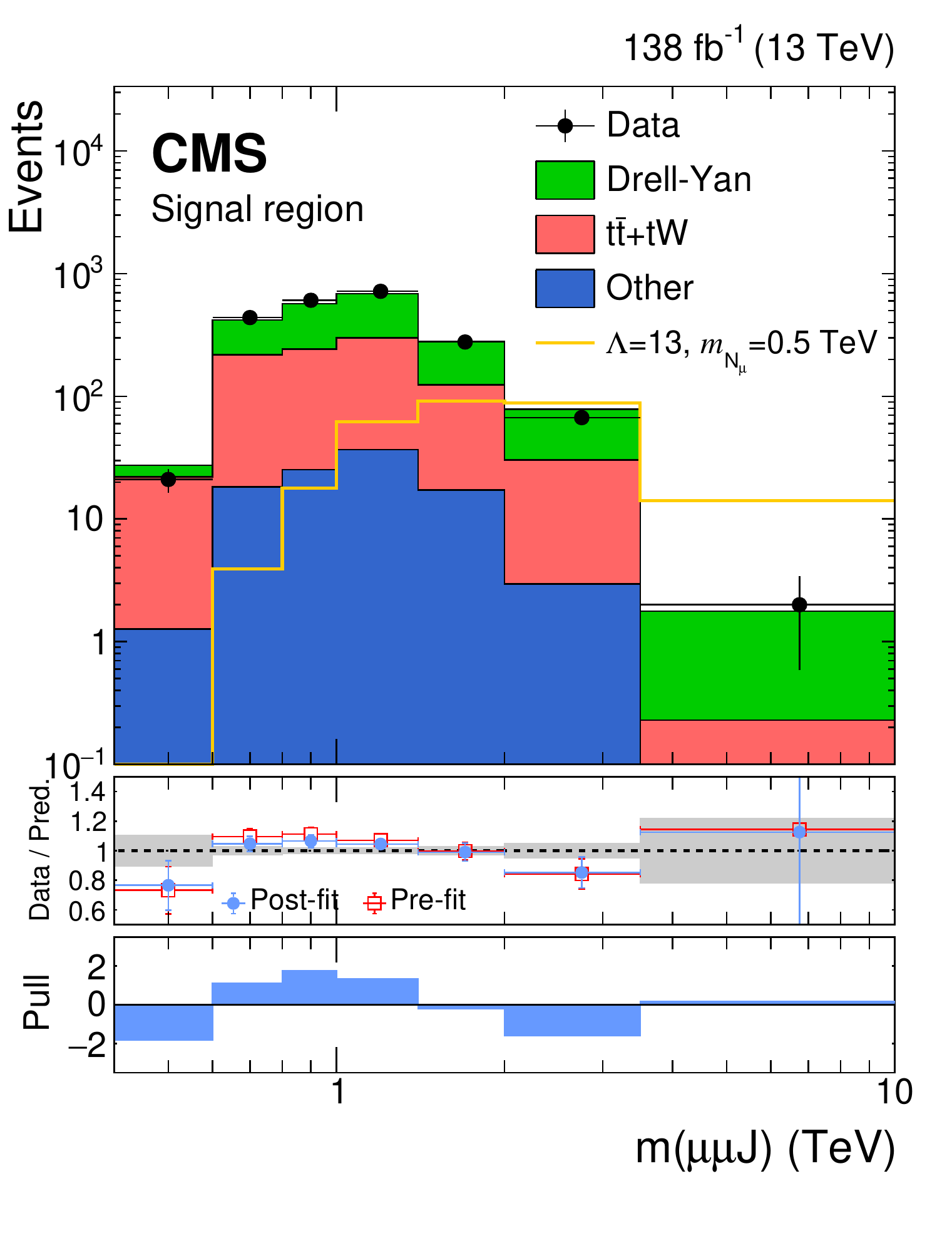}
\caption{Distributions of \mllj for the data, and the post-fit backgrounds (stacked histograms), in the SRs of the \eeqq (left) and the \mmqq (right) channels. 
The template for one signal hypothesis is shown overlaid as a yellow solid line.
The overflow is included in the last bin.
The middle panels show ratios of the data to the pre-fit background prediction and post-fit background yield as red open squares and blue points, respectively.
The gray band in the middle panels indicates the systematic component of the post-fit uncertainty.
The lower panels show the distributions of the pulls, defined in the text.
}
\label{fig:postfitSR}
\end{figure*}

From the ML fit we extract values of \sigstr for a range of the parameters \mhcmn and $\Lambda$ of the signal model.  The input data are the \mllj distribution in the SR and in the DY CR with $150 < \mll < 300\GeV$, and the \memuj distribution for the top quark-enriched CR (Fig.~\ref{fig:bkg CR}). 
The result of the fit under the background-only hypothesis is shown in Fig.~\ref{fig:postfitSR} for the \mllj distribution of the \eeqq and the \mmqq channels.

\begin{figure}[htb]
\centering
\includegraphics[width=0.49\textwidth]{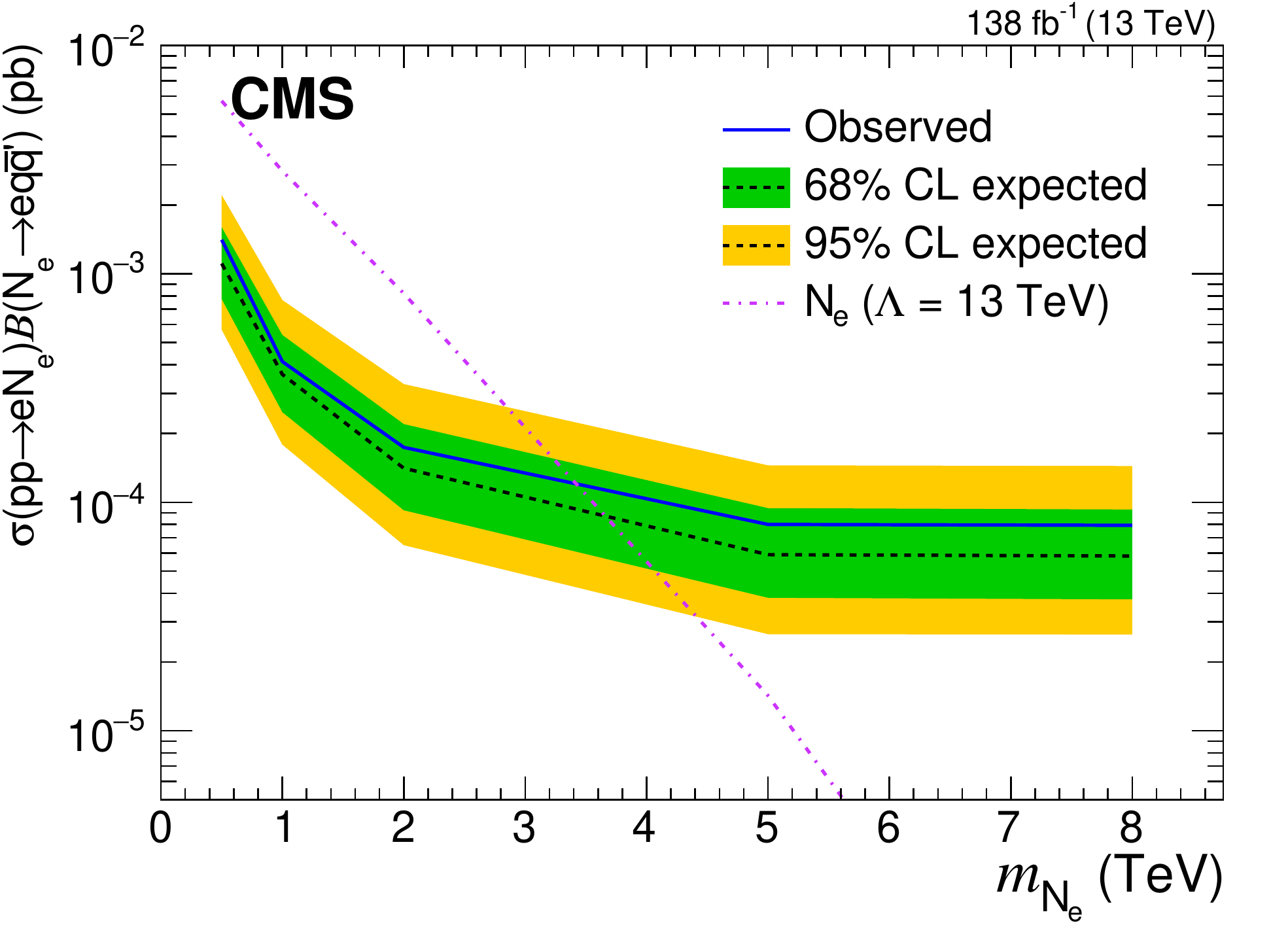}
\includegraphics[width=0.49\textwidth]{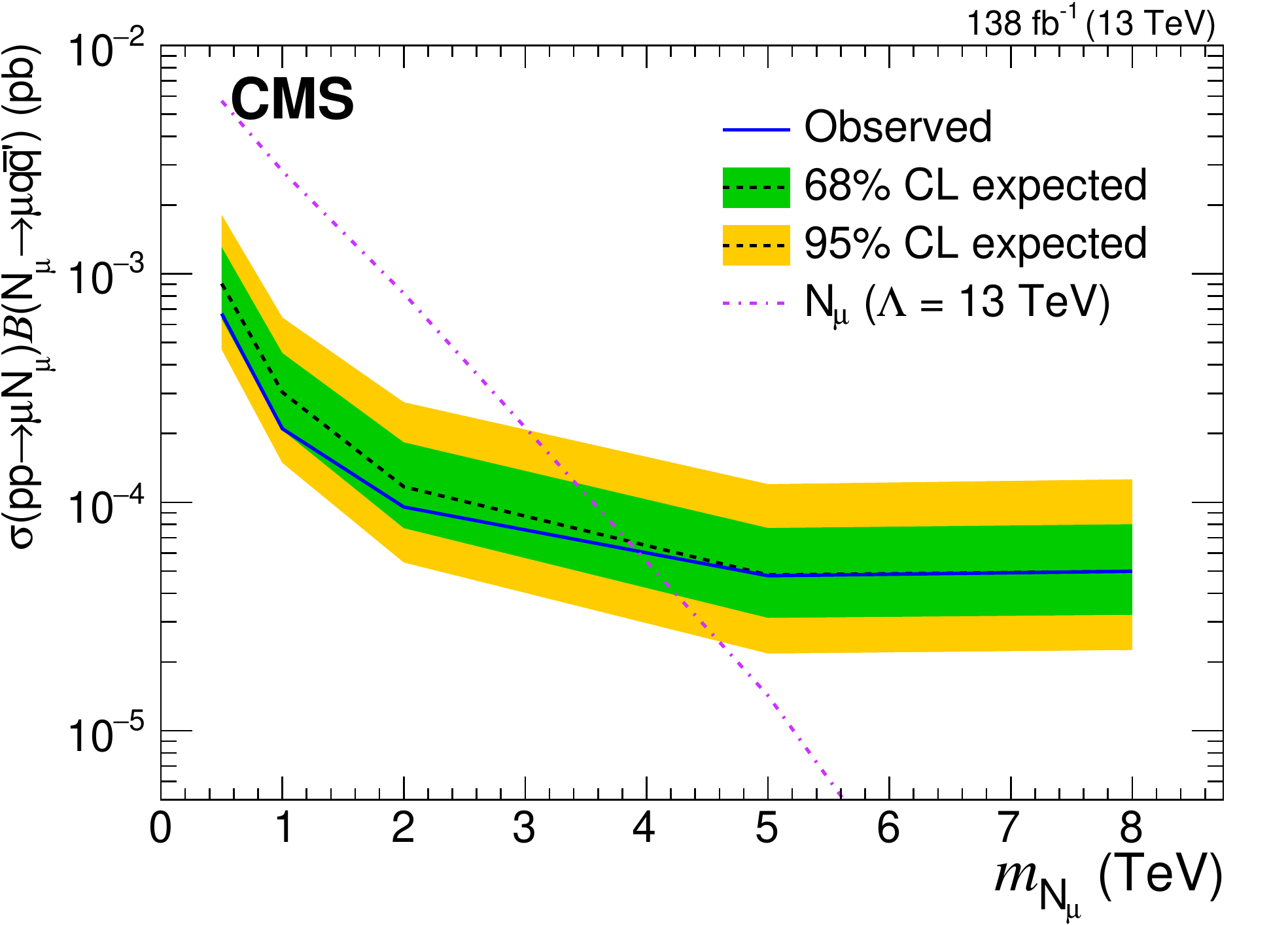}
\caption{Expected (black dashed lines with green dark and yellow light bands) and observed (solid blue lines) upper limits on the product of cross section and branching fraction for the \eeqq (\cmsLeft) and \mmqq (\cmsRight) channels. The uncertainty bands account for the post-fit statistical and systematic uncertainty. The magenta dot-dashed lines denote the model cross sections for the benchmark scale parameter $\Lambda = 13\TeV$.
}
\label{Limit_comb_with_toys}
\end{figure}

The observed data and the estimated SM background contributions are in agreement, and no significant excess is observed.  We derive upper limits at 95\% \CL on the product of cross section and branching fraction $\sigma(\Pp\Pp\to\ell\hcmnl)\mathcal{B}(\hcmnl\to\ell\qqbarp)$, using a \CLs method~\cite{Junk:1999kv,Read:2002hq}, in the asymptotic approximation~\cite{Cowan:2010js}.
The adequacy of the asymptotic approximation has been verified with pseudo-experiments.
The expected and observed upper limits for the \eeqq and the \mmqq channels are displayed in Fig.~\ref{Limit_comb_with_toys}, for a benchmark value of $\Lambda = 13\TeV$.  The limits are of order $10^{-4}\unit{pb}$ for a range of \hcmnl signal hypotheses.

\begin{figure}[htb]
\centering
\includegraphics[width=0.49\textwidth]{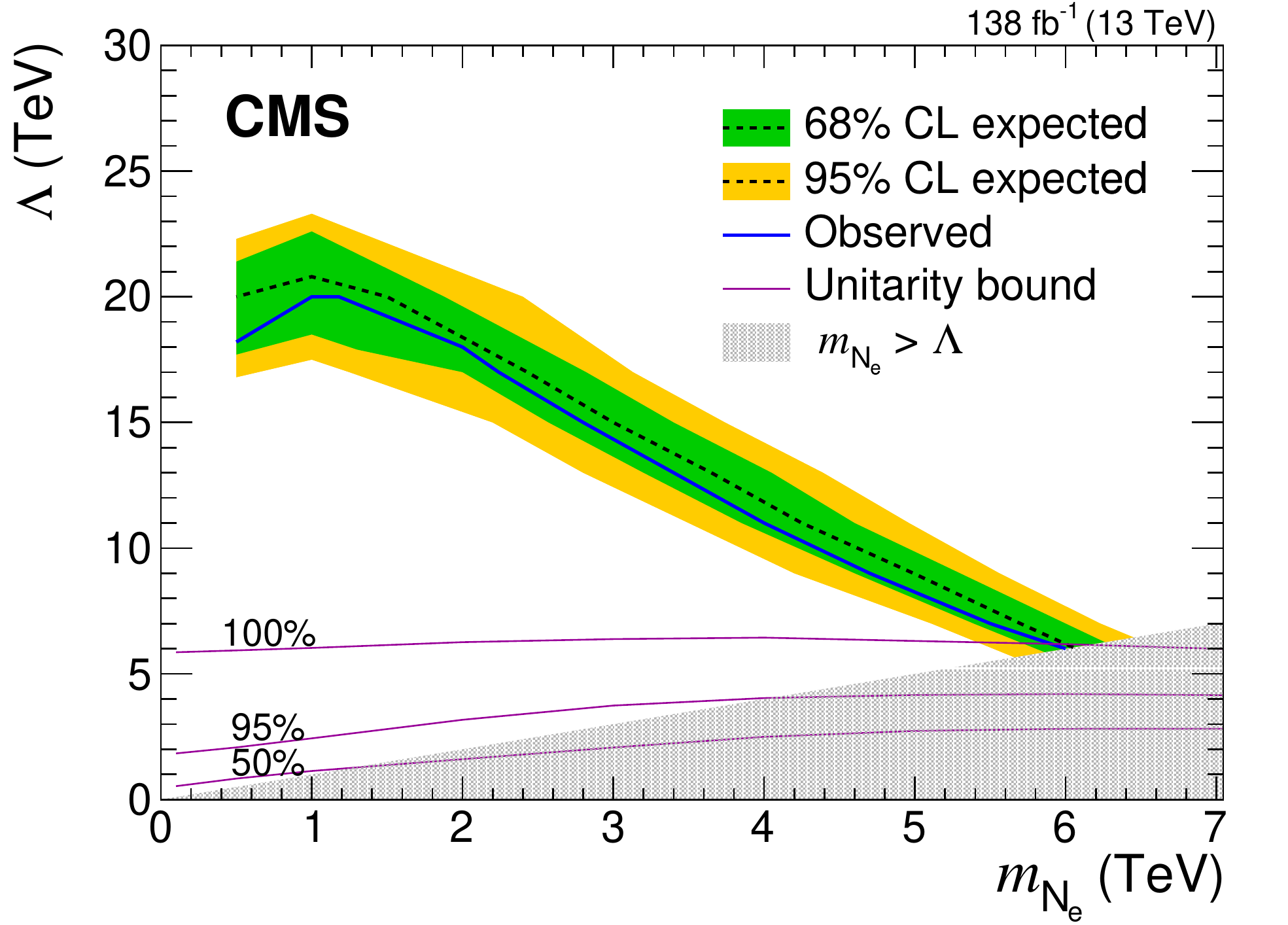}
\includegraphics[width=0.49\textwidth]{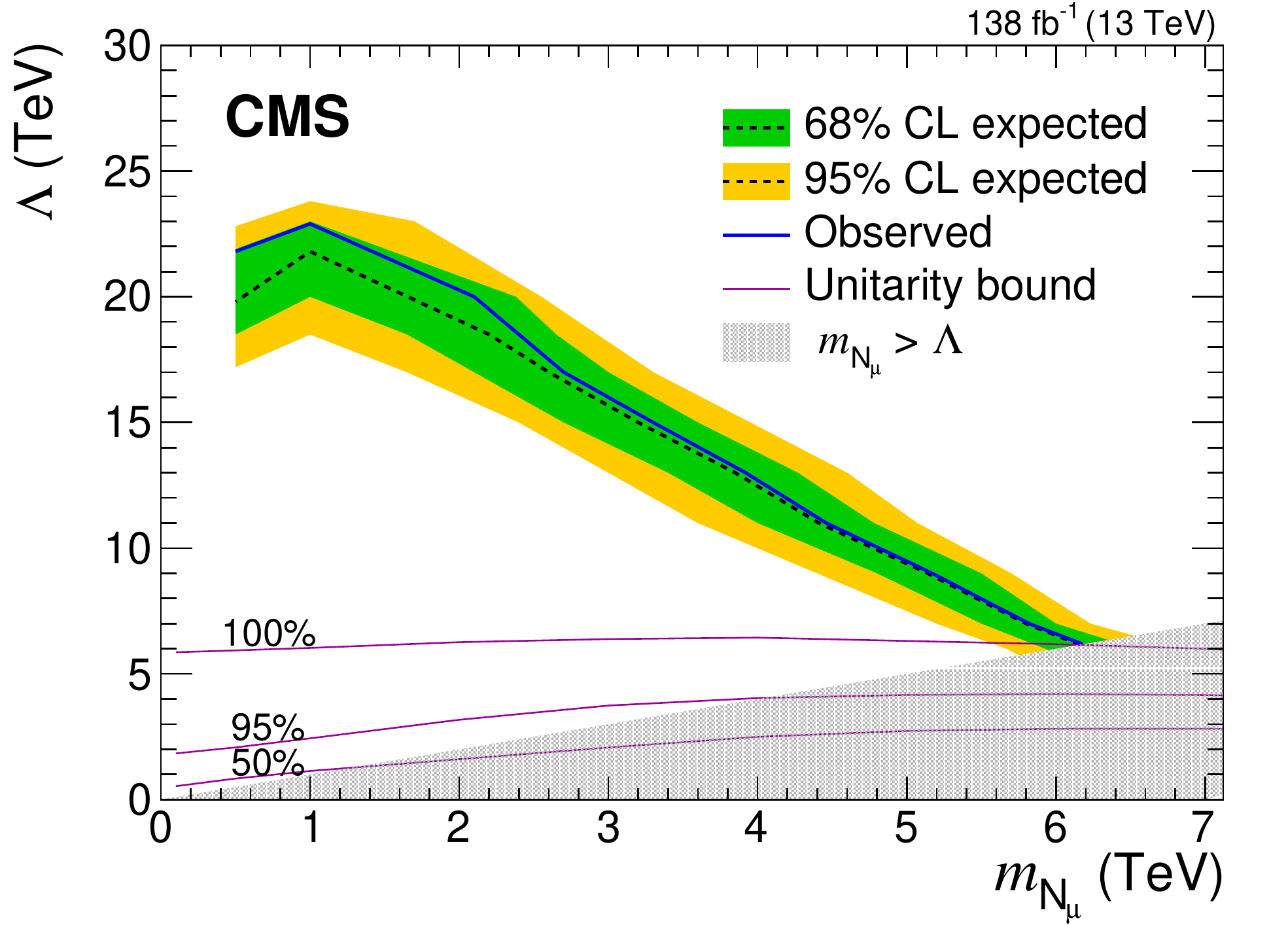}
\caption{Expected (black dashed lines with green dark and yellow light bands) and observed (solid blue lines) limits in the (\mhcmn, $\Lambda$) plane of the composite model for the \eeqq (\cmsLeft) and \mmqq (\cmsRight) channels. The region below the curve is excluded. The gray shading indicates the region where \mhcmn would exceed $\Lambda$, the EFT scale parameter, and the three solid magenta lines in the lower part of the plots represent the fraction of the signal-model phase space that satisfies the unitarity condition in the EFT approximation.
}
\label{2D_Limit_eejj}
\end{figure}

The results are recast in terms of the EFT of Ref.~\cite{Leonardi:2015qna} in Fig.~\ref{2D_Limit_eejj}, which shows the region in the (\mhcmn, $\Lambda$) plane that is excluded by the data. The region of validity is constrained by the model's assumption that $\mhcmn < \Lambda$.  A further consideration, discussed in Ref.~\cite{Biondini:2019tcc}, is that the unitarity of the scattering amplitude, as approximated in the perturbation expansion, can be violated for some values of the subenergies $\hat{s} \equiv s x_1 x_2$.  These subenergies appear in the integral over $x_{1,2}$ weighted by the product of proton PDFs $\mathcal{P}(x_1)\mathcal{P}(x_2)$. Contours giving the fraction of this ($x_1$, $x_2$) phase space that is consistent with unitarity are shown as the solid magenta curves in Fig.~\ref{2D_Limit_eejj}.
For the case of $\Lambda = \mhcmn$, the existence of \hcmne (\hcmnmu) is excluded by the data for masses up to 6.0 (6.1)\TeV at 95\% \CL, improving by more than 1\TeV the current most stringent limit on this class of resonances~\cite{2017315}, results that are safe from potential violation of the underlying EFT.
Moreover, the accessible range of $\Lambda$ is almost twice that reached in the previous search, extending the sensitivity to ${\approx}20\TeV$ at lower \mhcmn masses. 


\section{Summary}
\label{sec:summary}
A search is reported for a heavy composite Majorana neutrino \hcmnl, where the flavor $\ell$ corresponds to an electron or muon, that appears in composite fermion models.
In the specific model considered, the \hcmnl is produced in association with a lepton and subsequently decays into a same-flavor lepton plus two quarks, leading to a signature with two same-flavor leptons and at least one large-radius jet. The analysis is performed using a sample of proton-proton collisions at $\sqrt{s} = 13\TeV$ recorded by the CMS experiment at the CERN LHC, corresponding to an integrated luminosity of 138\fbinv.
The data are found to be in agreement with the standard model expectations.
In the context of an effective field theory with compositeness scale parameter $\Lambda$, an upper limit at 95\% \CL is established on $\sigma(\Pp\Pp\to\ell\hcmnl)\mathcal{B}(\hcmnl\to\ell\qqbarp)$ as a function of $\Lambda$ and the \hcmnl mass \mhcmn.
Masses less than 6.0 (6.1)\TeV are excluded for $\ell =\Pe\,(\PGm)$, at the limit $\mhcmn = \Lambda$.
For $\mhcmn \approx 1\TeV$, values of $\Lambda$ less than 20 (23)\TeV are excluded.
The present search covers a parameter space larger by about $1.6\TeV$ in \mhcmn compared to previous searches in proton-proton collisions at 13\TeV.

\begin{acknowledgments}
We congratulate our colleagues in the CERN accelerator departments for the excellent performance of the LHC and thank the technical and administrative staffs at CERN and at other CMS institutes for their contributions to the success of the CMS effort. In addition, we gratefully acknowledge the computing centers and personnel of the Worldwide LHC Computing Grid and other centers for delivering so effectively the computing infrastructure essential to our analyses. Finally, we acknowledge the enduring support for the construction and operation of the LHC, the CMS detector, and the supporting computing infrastructure provided by the following funding agencies: BMBWF and FWF (Austria); FNRS and FWO (Belgium); CNPq, CAPES, FAPERJ, FAPERGS, and FAPESP (Brazil); MES and BNSF (Bulgaria); CERN; CAS, MoST, and NSFC (China); MINCIENCIAS (Colombia); MSES and CSF (Croatia); RIF (Cyprus); SENESCYT (Ecuador); MoER, ERC PUT and ERDF (Estonia); Academy of Finland, MEC, and HIP (Finland); CEA and CNRS/IN2P3 (France); BMBF, DFG, and HGF (Germany); GSRI (Greece); NKFIH (Hungary); DAE and DST (India); IPM (Iran); SFI (Ireland); INFN (Italy); MSIP and NRF (Republic of Korea); MES (Latvia); LAS (Lithuania); MOE and UM (Malaysia); BUAP, CINVESTAV, CONACYT, LNS, SEP, and UASLP-FAI (Mexico); MOS (Montenegro); MBIE (New Zealand); PAEC (Pakistan); MES and NSC (Poland); FCT (Portugal); MESTD (Serbia); MCIN/AEI and PCTI (Spain); MOSTR (Sri Lanka); Swiss Funding Agencies (Switzerland); MST (Taipei); MHESI and NSTDA (Thailand); TUBITAK and TENMAK (Turkey); NASU (Ukraine); STFC (United Kingdom); DOE and NSF (USA).

\hyphenation{Rachada-pisek} Individuals have received support from the Marie-Curie program and the European Research Council and Horizon 2020 Grant, contract Nos.\ 675440, 724704, 752730, 758316, 765710, 824093, 884104, and COST Action CA16108 (European Union); the Leventis Foundation; the Alfred P.\ Sloan Foundation; the Alexander von Humboldt Foundation; the Belgian Federal Science Policy Office; the Fonds pour la Formation \`a la Recherche dans l'Industrie et dans l'Agriculture (FRIA-Belgium); the Agentschap voor Innovatie door Wetenschap en Technologie (IWT-Belgium); the F.R.S.-FNRS and FWO (Belgium) under the ``Excellence of Science -- EOS" -- be.h project n.\ 30820817; the Beijing Municipal Science \& Technology Commission, No. Z191100007219010; the Ministry of Education, Youth and Sports (MEYS) of the Czech Republic; the Hellenic Foundation for Research and Innovation (HFRI), Project Number 2288 (Greece); the Deutsche Forschungsgemeinschaft (DFG), under Germany's Excellence Strategy -- EXC 2121 ``Quantum Universe" -- 390833306, and under project number 400140256 - GRK2497; the Hungarian Academy of Sciences, the New National Excellence Program - \'UNKP, the NKFIH research grants K 124845, K 124850, K 128713, K 128786, K 129058, K 131991, K 133046, K 138136, K 143460, K 143477, 2020-2.2.1-ED-2021-00181, and TKP2021-NKTA-64 (Hungary); the Council of Science and Industrial Research, India; the Latvian Council of Science; the Ministry of Education and Science, project no. 2022/WK/14, and the National Science Center, contracts Opus 2021/41/B/ST2/01369 and 2021/43/B/ST2/01552 (Poland); the Funda\c{c}\~ao para a Ci\^encia e a Tecnologia, grant CEECIND/01334/2018 (Portugal); the National Priorities Research Program by Qatar National Research Fund;  MCIN/AEI/10.13039/501100011033, ERDF ``a way of making Europe", and the Programa Estatal de Fomento de la Investigaci{\'o}n Cient{\'i}fica y T{\'e}cnica de Excelencia Mar\'{\i}a de Maeztu, grant MDM-2017-0765 and Programa Severo Ochoa del Principado de Asturias (Spain); the Chulalongkorn Academic into Its 2nd Century Project Advancement Project, and the National Science, Research and Innovation Fund via the Program Management Unit for Human Resources \& Institutional Development, Research and Innovation, grant B05F650021 (Thailand); the Kavli Foundation; the Nvidia Corporation; the SuperMicro Corporation; the Welch Foundation, contract C-1845; and the Weston Havens Foundation (USA).    
\end{acknowledgments}

\bibliography{auto_generated}

\cleardoublepage \appendix\section{The CMS Collaboration \label{app:collab}}\begin{sloppypar}\hyphenpenalty=5000\widowpenalty=500\clubpenalty=5000
\cmsinstitute{Yerevan Physics Institute, Yerevan, Armenia}
{\tolerance=6000
A.~Tumasyan\cmsAuthorMark{1}\cmsorcid{0009-0000-0684-6742}
\par}
\cmsinstitute{Institut f\"{u}r Hochenergiephysik, Vienna, Austria}
{\tolerance=6000
W.~Adam\cmsorcid{0000-0001-9099-4341}, J.W.~Andrejkovic, T.~Bergauer\cmsorcid{0000-0002-5786-0293}, S.~Chatterjee\cmsorcid{0000-0003-2660-0349}, K.~Damanakis\cmsorcid{0000-0001-5389-2872}, M.~Dragicevic\cmsorcid{0000-0003-1967-6783}, A.~Escalante~Del~Valle\cmsorcid{0000-0002-9702-6359}, P.S.~Hussain\cmsorcid{0000-0002-4825-5278}, M.~Jeitler\cmsAuthorMark{2}\cmsorcid{0000-0002-5141-9560}, N.~Krammer\cmsorcid{0000-0002-0548-0985}, L.~Lechner\cmsorcid{0000-0002-3065-1141}, D.~Liko\cmsorcid{0000-0002-3380-473X}, I.~Mikulec\cmsorcid{0000-0003-0385-2746}, P.~Paulitsch, F.M.~Pitters, J.~Schieck\cmsAuthorMark{2}\cmsorcid{0000-0002-1058-8093}, R.~Sch\"{o}fbeck\cmsorcid{0000-0002-2332-8784}, D.~Schwarz\cmsorcid{0000-0002-3821-7331}, S.~Templ\cmsorcid{0000-0003-3137-5692}, W.~Waltenberger\cmsorcid{0000-0002-6215-7228}, C.-E.~Wulz\cmsAuthorMark{2}\cmsorcid{0000-0001-9226-5812}
% ...
\par}
\vskip\cmsinstskip
\dag:~Deceased\\
$^{1}$Also at Yerevan State University, Yerevan, Armenia\\
$^{2}$Also at TU Wien, Vienna, Austria\\
% ...
\end{sloppypar}
\end{document}